\documentclass[superscriptaddress,altaffilletter,amsmath,amssymb,aps,twocolumn,floatfix]{revtex4-1}

\usepackage[english]{babel}
\usepackage[colorinlistoftodos]{todonotes}
\usepackage{multirow}
\usepackage{comment}
\usepackage{relsize}
\usepackage{float}
\usepackage{textcomp}
\usepackage{hyperref}
\usepackage[justification=justified]{subcaption}
\usepackage[justification=justified, figurename=FIG.]{caption}
\usepackage[compact]{titlesec}
\titlespacing{\section}{0pt}{3ex}{2ex}
\titlespacing{\subsection}{0pt}{2ex}{2ex}
\titlespacing{\subsubsection}{0pt}{1.5ex}{1.5ex}

\DeclareCaptionLabelFormat{UC}{\MakeUppercase{#1} #2}
\usepackage{xcolor}
\usepackage{graphicx}
\usepackage{epstopdf}
\usepackage{color}
\usepackage{dcolumn}
\usepackage{bm}
\usepackage{hyperref}
\usepackage[mathlines]{lineno}
\usepackage{threeparttable} 
\usepackage{booktabs} 
\usepackage{siunitx}
\usepackage{placeins}
\usepackage{xspace}
\usepackage{tabularx}
\usepackage{adjustbox}

\newcommand{\Rnttt}{\ensuremath{^{222}\textrm{Rn}}\xspace} 
\newcommand{\Rnttz}{\ensuremath{^{220}\textrm{Rn}}\xspace} 
\newcommand{\Pbtof}{\ensuremath{^{214}\textrm{Pb}}\xspace} 
\newcommand{\Pbtot}{\ensuremath{^{212}\textrm{Pb}}\xspace} 
\newcommand{\KreF}{\ensuremath{^{85}\textrm{Kr}}\xspace} 

\newcommand\Tstrut{\rule{0pt}{2.6ex}}         

\begin{document}

\title{Improved calculations of beta decay backgrounds to new physics in liquid xenon detectors}

\author{S.J.~Haselschwardt}\email[Corresponding author: ]{scotthaselschwardt@lbl.gov}
\affiliation{Lawrence Berkeley National Laboratory, 1 Cyclotron Road, Berkeley, CA 94720, USA}

\author{J.~Kostensalo}
\affiliation{ University of Jyvaskyla, Department of Physics, P.O. Box 35, FI-40014, Finland}

\author{X.~Mougeot}
\affiliation{CEA, LIST, Laboratoire National Henri Becquerel, CEA-Saclay 91191 Gif-sur-Yvette Cedex, France}

\author{J.~Suhonen} 
\affiliation{ University of Jyvaskyla, Department of Physics, P.O. Box 35, FI-40014, Finland}

\date{\today}
\begin{abstract}
\noindent
We present high-precision theoretical predictions for the electron energy spectra for the ground-state to ground-state $\beta$ decays of $^{214}$Pb, $^{212}$Pb, and $^{85}$Kr most relevant to the background of liquid xenon dark matter detectors. The effects of nuclear structure on the spectral shapes are taken into account using large-scale shell model calculations. Final spectra also include atomic screening and exchange effects. The impact of nuclear structure effects on the $^{214}$Pb and $^{212}$Pb spectra below $\approx100$ keV, pertinent for several searches for new physics, are found to be comparatively larger than those from the atomic effects alone. We find that the full calculation for $^{214}$Pb ($^{212}$Pb) predicts 15.0-23.2\% (12.1-19.0\%) less event rate in a 1-15 keV energy region of interest compared to the spectrum calculated as an allowed transition when using values of the weak axial vector coupling in the range $g_{\rm A}=0.7-1.0$. The discrepancy highlights the importance of both a proper theoretical treatment and the need for direct measurements of these spectra for a thorough understanding of $\beta$ decay backgrounds in future experiments.
\end{abstract}

\maketitle

\section{\label{sec:intro}INTRODUCTION}

The discovery of rare events caused by new physics requires that backgrounds which could mimic the signal be reduced as much as possible. Irreducible backgrounds must be well studied with credible estimates of their uncertainties. Searches for new physics based on the dual-phase liquid xenon (LXe) time projection chamber (TPC) have exciting potential for the discovery of dark matter and for the discovery of new neutrino properties. In the last decade these experiments have grown in size while background levels have been suppressed.

In current and future LXe TPCs, the majority of background rate in the low-energy ($\lesssim50$~keV) regime results from $\beta$ decays of \Pbtof, \Pbtot, and \KreF, with \Pbtof being the most significant of these by far. The isotopes \Pbtof and \Pbtot enter the LXe bulk as daughters of \Rnttt and \Rnttz which emanate out of the detector construction materials and dust. \KreF, on the other hand, enters through its abundance in the atmosphere and is therefore present in the raw xenon feedstock. The residual quantity found in low-background LXe experiments is that which survives xenon purification techniques such as chromatography~\cite{Akerib:2016hcd} and distillation~\cite{Aprile:2016xhi}. Typically, the background from \Pbtof dominates over that from \Pbtot and \KreF owing to the \Rnttt half-life of 3.8~days.

Each of these isotopes exhibit a $\beta$ decay in which the transition proceeds directly to the daughter nuclei's ground state with no associated $\gamma$-ray or conversion electron emission (henceforth referred to as the ``ground-state" decay). The result is a continuous energy distribution of single-site events made by the ejected electron which spans from zero up to the decay $Q$-value. Table~\ref{tab:data} provides a summary of pertinent nuclear data for these three isotopes and their ground state transitions. The low energy population of these decays forms the majority background for many new physics searches as illustrated in Table~\ref{tab:bkg}. The remaining $\beta$ decays which populate excited states of the daughter and result in $\gamma$-ray emission are less of a concern as they are more easily identified as multi-site events which do not mimic the sought-after signal. While multiple techniques are used to infer the final level of each isotope realized in an experiment, the modelling of this background depends on the ground state decay branching ratios assumed and, more crucially, the precise shape of the $\beta$ energy spectra.

\begin{table}[t]
\centering
\caption{ Relevant nuclear data for the isotopes considered in this work. The last three columns provide information pertaining to ground state decays, including the branching ratio (BR) and the initial and final spin-parity assignments, $J_i^{\pi}$ and $J_f^{\pi}$. Uncertainties smaller than 5\% are not shown. Endpoint data is from~\cite{Wan17} and all other data is from~\cite{A214datasheet,A212datasheet,A85datasheet}. }
\label{tab:data}
\begin{ruledtabular}
\begin{tabular}{llccc}
\multirow{2}{*}{Isotope} & \multirow{2}{*}{Half-life} & \multicolumn{3}{c}{Ground state $\beta$ decay}  \\
 & & Endpoint (keV) & BR (\%) & $J_i^{\pi},J_f^{\pi}$ \\
\hline
\Pbtof & 26.8~min & 1018 & 11.0(10)\Tstrut & $0^+,1^-$ \\
\Pbtot & 10.6~h & 569.1 & 11.9(16) & $0^+,1^-$ \\
\KreF & 10.7~yr & 687.0 & 99.6 & $9/2^+,5/2^-$ \\
\end{tabular}
\end{ruledtabular}
\end{table}

\begin{table}[ht]
\centering
\caption{ Projected and measured percentage of total electron recoil background in LXe TPC experiments attributed to the ground-state $\beta$ decays of the given isotopes in the specified energy windows. Other backgrounds arise from solar neutrino-electron scattering, $2\nu\beta\beta$ decay of $^{136}$Xe, and $\gamma$-rays from detector materials. The contribution from $^{136}$Xe is falling steeply in this region, becoming subdominant to that from solar neutrino scattering below $\approx12$~keV. }
\label{tab:bkg}
\begin{ruledtabular}
\begin{tabular}{lccc}
\multirow{2}{*}{Isotope} & LZ~\cite{LZ:2018} & XENONnT~\cite{xntSens} & XENON1T~\cite{Aprile:2020tmw} \\
 & 1.5--15~keV & 1--13~keV & 1--30~keV \\
\hline
\Pbtof & 53 & 42 & $83\pm2$ \Tstrut\\
\Pbtot & 8.8 & - & - \\
\KreF & 2.3 & 8.2 & $10\pm2$ \\
\end{tabular}
\end{ruledtabular}
\end{table}

The shape of a $\beta$ particle energy spectrum depends on the nature of the weak interaction transition and on both the atomic and nuclear structure of the initial and final states involved. For first-forbidden unique decays, such as the decay of \KreF, there are only small corrections to the spectrum shape from nuclear structure. However, for first-forbidden non-unique transitions, such as the ground state decays of \Pbtof and \Pbtot, the spectral shape can depend heavily on the details of the nuclear structure.

The presently used formalism for the forbidden non-unique $\beta$ transitions was first introduced in~\cite{Mustonen2006} and later extended in~\cite{Haaranen2016} and~\cite{Haaranen2017} to include the next-to-leading-order corrections to the $\beta$-decay shape function. In~\cite{Haaranen2016} it was noticed for the first time that some of the forbidden non-unique $\beta$ transitions can depend strongly on the effective value of the weak axial vector coupling constant $g_{\rm A}$. This dependence on $g_{\rm A}$ was studied in the nuclear shell-model framework in~\cite{Kostensalo2017}. A recent review of the $\beta$ spectral-shape calculations is given in~\cite{Ejiri2019}. The present shell-model calculations are an extension of the aforementioned $\beta$-decay formalism in that the magnitude of a key vector-type nuclear matrix element (NME) is fixed to reproduce the partial half-life of the ground state transitions of \Pbtot and \Pbtof. This method was used in recent $\beta$-decay calculations for light nuclei in~\cite{Kumar2020}.

Recently, the XENON1T experiment reported~\cite{Aprile:2020tmw} an excess of electron recoil events above a background which is dominated by the ground-state $\beta$ decay of \Pbtof. In that result both the nuclear transition and atomic exchange effects were modelled assuming the decay is an allowed transition. In this work we report on the ground-state $\beta$ shapes of \Pbtof and \Pbtot obtained by calculating the necessary NMEs for a first-forbidden non-unique transition and by employing a formalism for atomic exchange corrections that has been extended to include forbidden unique transitions. The same exchange formalism is then also applied to the ground state $\beta$ decay of \KreF.

\section{\label{sec:calcs}CALCULATIONS}

The half-life of a forbidden non-unique $\beta^-$ decay can be expressed as
\begin{equation}
t_{1/2}=\tilde{\kappa}/\tilde{C},
\label{eq:hl}
\end{equation}
where~\cite{Hardy1990}

\begin{equation}
\tilde{\kappa} = \frac{2\pi^3\hbar^7\mathrm{ln \  2}}{m_e^5c^4(G_{\rm F}
\cos \theta_{\rm C})^2}= 6147 \ \mathrm{s},
\label{eq:kappa}
\end{equation}
$\theta_{\rm C}$ being the Cabibbo angle and $\tilde{C}$ is the dimensionless
integrated shape function, given by
\begin{equation}
\tilde{C} = \int^{w_0}_1 C(w_e)pw_e(w_0-w_e)^2F_0(Z,w_e)K(w_e)dw_e,
\label{eq:ctilde}
\end{equation} 
where $w_e=W_e/m_ec^2$, $w_0 = W_0/m_ec^2$, and 
$p=p_ec/(m_ec^2)= \sqrt{w_e^2 -1}$ are unitless kinematic 
quantities, $F_0(Z,w_e)$ is the Fermi function, and $K(w_e)$ encompasses a plurality of corrective terms such as atomic effects.
The shape factor $C(w_e)$ of Eq.~(\ref{eq:ctilde}) contains complicated combinations of
both (universal) kinematic factors and nuclear form factors. The nuclear form factors 
can be related to the corresponding NMEs using the impulse approximation~\cite{Behrens1982}. 

The $\beta$ particle spectrum is given by the integral in Eq.~(\ref{eq:ctilde}). The probability of the electron being emitted with energy between $w_e$ and $w_e+dw_e$ is
\begin{equation}
    P(w_e)dw_e \propto C(w_e)pw_e(w_0-w_e)^2F_0(Z,w_e)K(w_e)dw_e.
\end{equation}

\subsection{\label{ssec:nuclear_calcs}Nuclear shape factors}

For the first-forbidden decays the relevant NMEs are those corresponding to the transition operators
\begin{align}
\mathcal{O} (0^-): g_{\rm A}({\bm\sigma \cdot {\textbf{p}}_e}), 
\quad g_{\rm A} &({\bm\sigma \cdot {\textbf{r}}}) \label{eq:rank-0}\\
\mathcal{O} (1^-): g_{\rm V}\textbf{p}_e, \quad g_{\rm A} ({\bm\sigma 
\times {\textbf{r}}}), &\quad g_{\rm V}\textbf{r}  \label{eq:rank-1} \\
\mathcal{O} (2^-): g_{\rm A} [{\bm\sigma {\textbf{r}}}]_2&, \label{eq:rank-2}
\end{align}
where \textbf{r} is the radial coordinate and $\textbf{p}_e$ is the electron momentum. The decay of \KreF is first-forbidden unique, so only the operator $ g_{\rm A} [{\bm\sigma {\textbf{r}}}]_2$ contributes, simplifying the calculations greatly. For the ground state decay of \Pbtot and \Pbtof only the rank-1 operators contribute. The NMEs involved in the transitions can be evaluated using the relation
\begin{align}
\begin{split}
^{V/A}\mathcal{M}_{KLS}^{(N)}(pn)(k_e,m,n,\rho)& \\=\frac{\sqrt{4\pi}}{\widehat{J}_i}
\sum_{pn} \, ^{V/A}m_{KLS}^{(N)}(pn)(&k_e,m,n,\rho)(\Psi_f|| [c_p^{\dagger}
\tilde{c}_n]_K || \Psi_i),
\label{eq:ME}
\end{split}
\end{align} 
where $^{V/A}m_{KLS}^{(N)}(pn)(k_e,m,n,\rho)$ is the single-particle matrix element, and $(\Psi_f|| [c_p^{\dagger}\tilde{c}_n]_K || \Psi_i)$ is the one-body transition density (OBTD), which contains all the relevant nuclear-structure information. The OBTDs need to be evaluated using some nuclear model, such as the nuclear shell model used in this work. The nuclear structure calculations were done using the shell-model code NuShellX@MSU~\cite{nushellx}. For \KreF the calculations were carried out in the full $0f_{5/2}$--$1p$--$0g_{9/2}$ valence space with the effective Hamiltonian JUN45~\cite{Honma2009}. For the Pb isotopes the calculations were done using the complete valence space spanned by proton orbitals $0h_{9/2}$, $1f$, $2p$, and $0i_{13/2}$ and neutron orbitals $0i_{11/2}$, $1g$, $2d$, $3s$, and $0j_{15/2}$ with the effective Hamiltonian khpe~\cite{Warburton91}. 

For \KreF the spectral shape does not depend on the nuclear structure in the leading-order terms. In this work we include also the next-to-leading-order terms in the Behrens and B{\"u}hring expansion~\cite{Behrens1982}, which increases the number of NMEs involved in each transition to 5 for \KreF and to 13 for the non-unique transitions.

Uncertainties in the theoretical spectral shapes are related to uncertainties in the ratios of the NMEs. Based on previous calculations, quenching of the ratio  of the axial-vector and vector coupling constants $g_{\rm A}/g_{\rm V}$ is needed in order to reproduce experimental spectral shapes for non-unique beta decays with the shell model~\cite{Haaranen2017}. However, the precise amount of quenching needed for the decays studied here is not known. Based on previous studies from the past four decades, the value $g_{\rm A}=1.0$ was chosen for \KreF, while for $^{212,214}$Pb we report results using the range of values $g_{\rm A}=0.85\pm0.15$ as the quenching of $g_{\rm A}$ seems to be more severe for larger masses (see e.g.~\cite{Suhonen2017}). Since the decay of \KreF is first-forbidden unique, the value of $g_{\rm A}$ affects only the next-to-leading order terms resulting in a correction on the order $\sim$0.1\%. On the other hand, in non-unique decays the value of $g_{\rm A}$ can be more impactful and different values can result in different spectral shapes~\cite{Haaranen2017}. In $^{212,214}$Pb this is not the case, and here we find the spectral shapes are somewhat insensitive to the value of $g_{\rm A}$.

The experimental half-lives of the ground state $^{212,214}$Pb transitions are reproduced within the chosen range of $g_{\rm A}$. Specifically, this is achieved with the value 0.83 (0.91) for \Pbtof (\Pbtot) without quenching $g_{\rm V}$ from the conserved vector current hypothesis value of 1.0. The ratio $g_{\rm A}/g_{\rm V}$ also agrees with other shell model calculations in this mass region. Warburton's calculations in this mass region resulted in the value $g_{\rm A}/g_{\rm V}=0.6/0.6\approx 1.0$~\cite{Warburton91} and more recent calculations of Zhi \emph{et al.} in $g_{\rm A}/g_{\rm V}=0.48/0.65\approx 0.74$~\cite{Zhi2013}. It should be noted that the spectral shape is only sensitive to ratios of matrix elements, and so the absolute quenching factor of all matrix elements is irrelevant. Thus taking $g_{\rm A}=g_{\rm V}=0.6$ will result in the same spectral shape as $g_{\rm A}=g_{\rm V}=1.0$. For values of $g_{\rm A}$ which did not manage to reproduce the experimental half-life, the small matrix element $^V\mathcal{M}_{101}$ was adjusted so that the experimental partial half-life related to the transition was reproduced. This approach was shown to work well in the case of the second-forbidden non-unique decay of $^{36}$Cl in the recent work of Kumar \emph{et al.}~\cite{Kumar2020}.
 
Recently, the EXO-200 collaboration reported~\cite{EXO-200} a measurement of the $\beta$ shape of the first-forbidden non-unique ground state $\beta$ transition $^{137}\textrm{Xe}(7/2^-)\to\,^{137}\textrm{Cs}(7/2^+)$. Good agreement between the measured $\beta$ spectrum and that computed in our formalism was found. This transition is not entirely analogous to the decays considered here, however, as rank-0 matrix elements play a significant role. Furthermore, the change in the shape factor is small, about 2\%.

In the mass region of interest here, the decay of $^{210}$Bi could provide a useful comparison as it connects $1^-$ and $0^+$ states.
However, the spectral shape of $^{210}$Bi computed in \cite{kostensalo2018} was found to depend strongly on the adopted value of $g_{\rm A}$. We therefore do not consider a comparison with this transition a valid test of our calculations, as almost any spectral shape could be fit by altering $g_{\rm A}$. The transition in $^{210}$Bi furthermore differs from those in $^{212,214}$Pb because of the differences in nuclear structure discussed in the appendix of~\cite{Aprile:2020tmw}.

\subsection{\label{ssec:nuclear_calcs}Atomic exchange effect}

The exchange effect has already been demonstrated to be the most prominent atomic effect at low energy~\cite{Hay18}, possibly enhancing the decay probability by more than 10\% below 5~keV \cite{Harston92,Mou14}. It arises from the indistinguishability of the electrons and the imperfect orthogonality of the initial and final atomic states due to the change of the nuclear charge in the decay. The exchange process is an additional decay channel with the same final state as the direct decay and can be seen as the swap of the $\beta$ electron with an electron of the atomic cloud, which is then ejected to the continuum.

Previous studies that included this effect were only focused on allowed transitions or assumed that the correction for an allowed transition can be applied as a first approximation to a forbidden transition~\cite{Harston92,Mou14,Kossert15,Kossert18,Aprile:2020tmw}. This is because a precise formalism of the exchange effect was set out only for allowed transitions~\cite{Pyper88}. A summary of the key ingredients that are needed to calculate the exchange correction factor can be found in~\cite{Aprile:2020tmw}.

The $\beta$ spectra calculated as allowed in the present work are identical to the ``improved calculations'' in~\cite{Aprile:2020tmw}. Full numerical calculation of the atomic screening and exchange effects is included, as well as accurate radiative corrections from the precise study of superallowed transitions~\cite{Tow08}. Identical calculations have also been performed for the ground state \KreF decay but with the exchange effect correctly determined for this first-forbidden unique transition.

Indeed, the formalism from~\cite{Pyper88} has recently been extended to the forbidden unique transitions and will be detailed elsewhere. We briefly summarize the main results here. The definition of the relativistic electron wave functions is consistent with Behrens and B{\"u}hring formalism~\cite{Behrens1982}. In the relativistic case, the usual operator $L^2$ defined from the orbital angular momentum operator $\vec{L}$ does not commute with the Hamiltonian. Instead, the appropriate operator to consider is
\begin{equation}
\hat{K} = \beta(\vec{\sigma}\cdot\vec{L}+1) \, ,
\end{equation}
with $\beta$ the ($4 \times 4$) Dirac matrix and $\vec{\sigma}$ standing for the three ($4 \times 4$) matrices defined from the ($2 \times 2$) Pauli matrices $\sigma_{x,y,z}$. Its eigenvalue is the quantum number $\kappa$ and it is convenient to introduce the quantity $k = |\kappa|$. Under spherical symmetry, only the small and large radial components are of interest. A continuum state is characterized by its quantum number $\kappa$, its total energy $w_e$, its momentum $p$, and its Coulomb amplitude $\alpha_{\kappa}$, and is denoted $\phi_{c,\kappa}$. Similarly, an atomic bound state is characterized by its quantum numbers ($n$,$\kappa$), its binding energy $E_{n \kappa}$, its total energy $w_{n \kappa} = 1 - |E_{n \kappa}|/m_e c^2$, its momentum $p_{n \kappa} = \sqrt{1 - w_{n \kappa}^2}$, and its Coulomb amplitude $\beta_{n \kappa}$, and is denoted $\phi_{b,n \kappa}$. Primed quantities refer to the daughter atom and to the parent atom otherwise. 

By restricting to the dominant NMEs, $\beta$ electrons can only be created in states with $\kappa = \pm 1$ in allowed transitions. The exchange process can then occur only for the atomic electrons in $s_{1/2}$ ($\kappa = -1$) and $p_{1/2}$ ($\kappa = +1$) orbitals. The shape factor $C(w_e)$ as defined in Eq.~(\ref{eq:ctilde}) being energy independent, the exchange effect is corrected by applying 
\begin{equation}
\label{eq:excorr_all}
C(w_e) \text{ } \longrightarrow \text{ } C(w_e) \times (1+\eta_1) \, .
\end{equation}
In the case of first-forbidden unique transitions, $\beta$ electrons can be created in states with $\kappa = \pm 1\text{, } \pm 2$. The exchange process can thus occur also for the atomic electrons in $p_{3/2}$ ($\kappa = -2$) and $d_{3/2}$ ($\kappa = +2$) orbitals. In addition, the shape factor is well-known to exhibit the following energy dependence:
\begin{equation}
\label{eq:1fu_shape}
C(w_e) \propto q^2 + \lambda_2 p^2 \, ,
\end{equation}
with $q = (w_0 - w_e)$ and 
\begin{equation}
\lambda_2 = \dfrac{\alpha_{+2}^2 + \alpha_{-2}^2}{\alpha_{+1}^2 + \alpha_{-1}^2} \, .
\end{equation}
The first term in Eq.~(\ref{eq:1fu_shape}) comes from $\beta$ electrons with $\kappa = \pm 1$ and the second term from those with $\kappa = \pm 2$. One can demonstrate that the exchange effect is corrected by applying
\begin{equation}
\label{eq:excorr_1}
q^2 \text{ } \longrightarrow \text{ } q^2 \times (1+\eta_1)
\end{equation}
\begin{equation}
\label{eq:excorr_2}
\lambda_2 p^2 \text{ } \longrightarrow \text{ } \lambda_2 p^2 \times (1+\eta_2) \, .
\end{equation}
The correction factor is defined by
\begin{equation}
\eta_k = \dfrac{T_{+k}(T_{+k} - 2\alpha_{+k}) + T_{-k}(T_{-k} - 2\alpha_{-k})}{\alpha_{+k}^2 + \alpha_{-k}^2} \, .
\end{equation}
The exchange probability between a $\beta$ electron and an atomic electron mainly depends on the overlap of their radial wave functions. As the process can occur with each electron in a $\kappa$ state, one has to sum over the different ($n$,$\kappa$) states. Assuming no atomic excitation and completely filled orbitals, one can establish
\begin{equation}
T_{\kappa} = \sum\limits_{n}{\text{ } \dfrac{\langle\phi_{c,\kappa}'|\phi_{b,n \kappa}\rangle}{\langle\phi_{b,n \kappa}'|\phi_{b,n \kappa}\rangle} \text{ } \beta_{n \kappa}' \text{ } \left( \dfrac{p_{n \kappa}'}{p} \right)^{k-1}} \, .
\end{equation}

In the case of allowed transitions, this result is similar to what is described in~\cite{Pyper88} except that the overlap of parent and daughter atomic wave functions is no longer approximated by unity. As in~\cite{Aprile:2020tmw}, the relativistic electron wave functions have been determined following the numerical procedure described in~\cite{Mou14}, forcing the convergence to the accurate orbital energies from~\cite{Kot97} for the bound states.

\section{\label{sec:results}RESULTS}

\begin{figure*}[b]
\centering
\begin{subfigure}{0.49\textwidth}
\includegraphics[width=\textwidth]{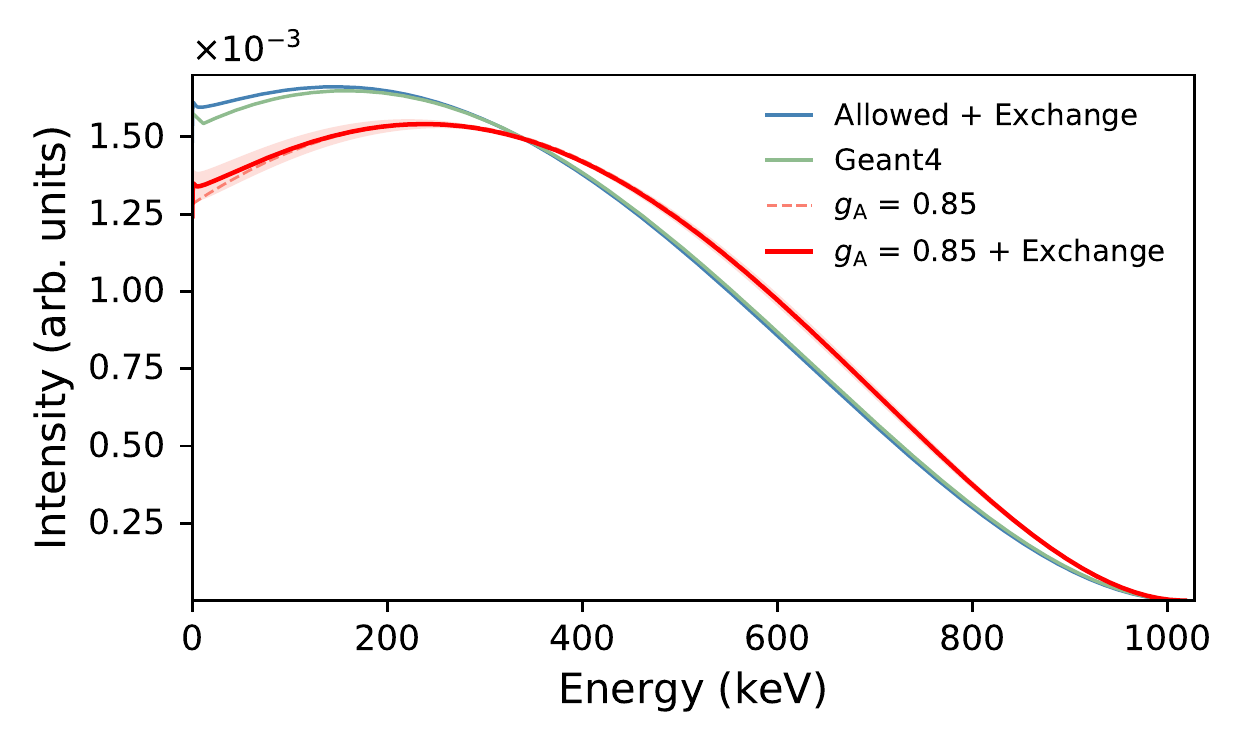}
\end{subfigure}
\begin{subfigure}{0.49\textwidth}
\includegraphics[width=\textwidth]{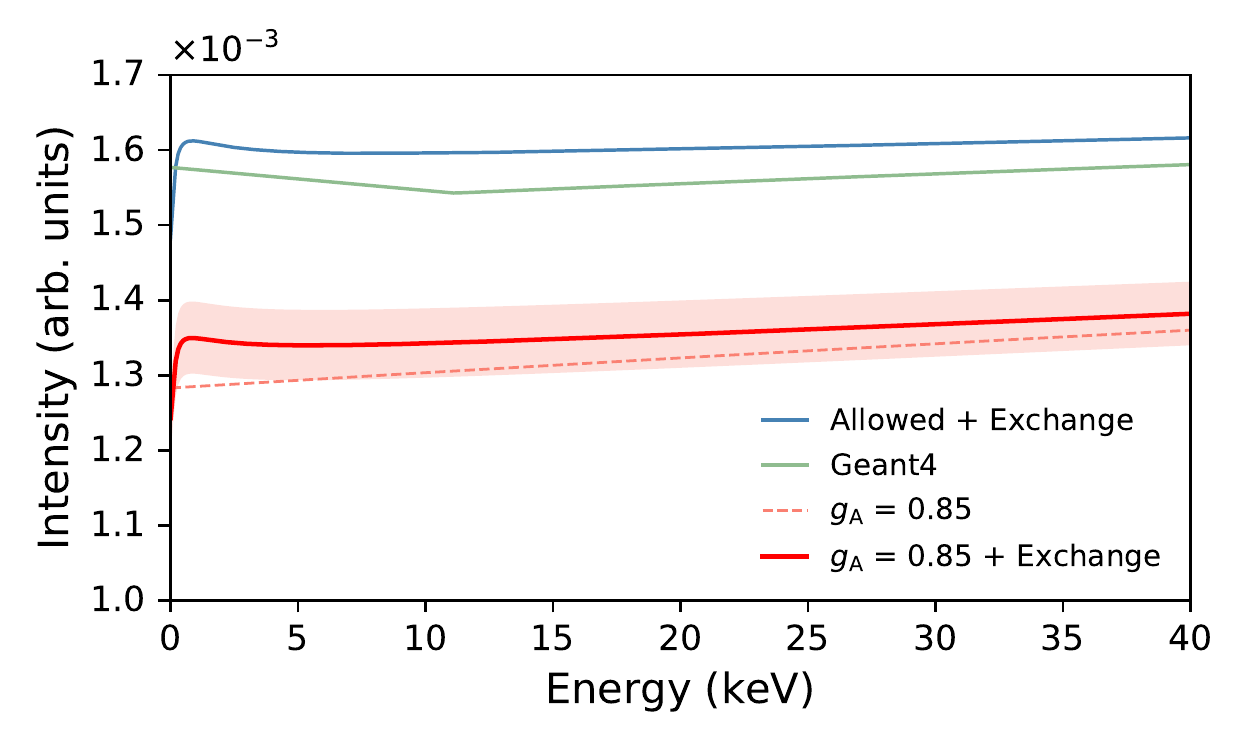}
\end{subfigure}
\caption{Comparison of $\beta$ spectra for the ground-state decay of \Pbtof shown over the full energy range (left) and at low energies (right). The result of this work is shown as a solid red line calculated using $g_{\rm A}=0.85$ and applying the atomic exchange correction. The upper and lower bounds of the shaded band show the spectrum obtained with $g_{\rm A}=0.7$ and $g_{\rm A}=1.0$, respectively. Spectra are normalized over the full energy range for each value of $g_{\rm A}$. \label{fig:pb214}}
\end{figure*}

\begin{figure*}[htb]
\centering
\begin{subfigure}{0.49\textwidth}
\includegraphics[width=\textwidth]{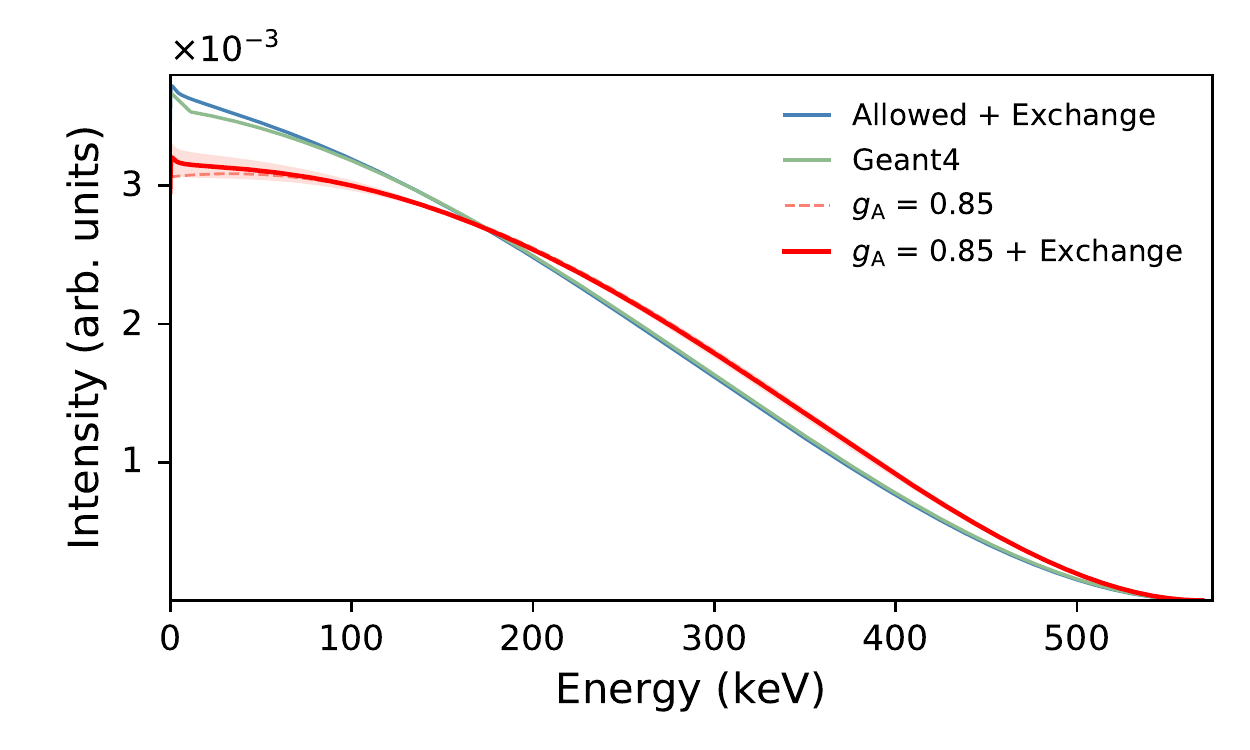}
\end{subfigure}
\begin{subfigure}{0.49\textwidth}
\includegraphics[width=\textwidth]{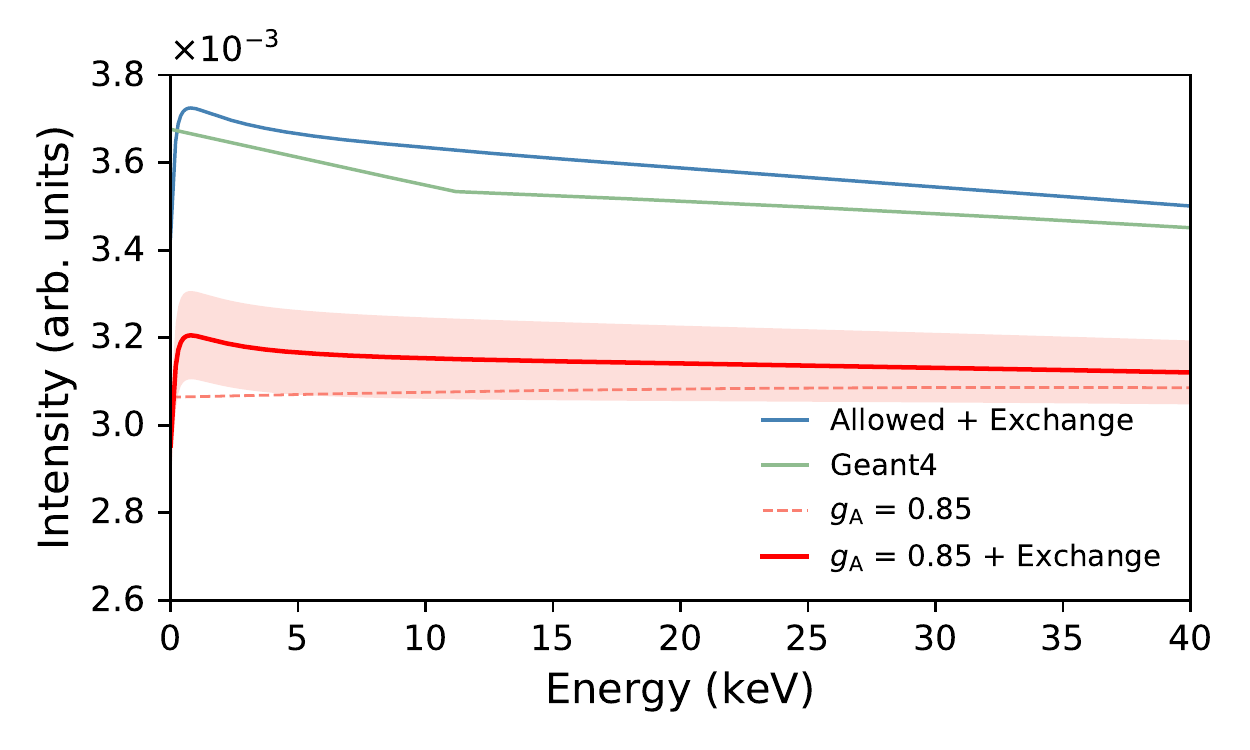}
\end{subfigure}
\caption{Comparison of $\beta$ spectra for the ground-state decay of \Pbtot shown over the full energy range (left) and at low energies (right). The result of this work is shown as a solid line calculated using $g_{\rm A}=0.85$ and applying the atomic exchange correction. The upper and lower bounds of the shaded band show the spectrum obtained with $g_{\rm A}=0.7$ and $g_{\rm A}=1.0$, respectively. Spectra are normalized over the full energy range for each value of $g_{\rm A}$. \label{fig:pb212}}
\end{figure*}

The final ground-state $\beta$ spectra of \Pbtof and \Pbtot obtained from our calculations are shown in Figs.~\ref{fig:pb214} and \ref{fig:pb212}, respectively. In each figure, the final spectrum shown in solid red includes both the effects of nuclear structure and the atomic exchange effect and is evaluated using the value of $g_{\rm A}=0.85$. The spectrum without the exchange correction (nuclear structure only) is shown as a dashed line for comparison. Surrounding the final spectrum is a band which shows the impact of varying the value of $g_{\rm A}$ from 0.7 to 1.0, and these values comprise the upper- and lower-band boundaries, respectively. The boundaries have been calculated using the same normalization as the solid line. The differences in the low energy part of the spectrum are therefore due to the change in the ratios of the relevant matrix elements rather than half-life.

\begin{figure*}[ht]
\centering
\begin{subfigure}{0.49\textwidth}
\includegraphics[width=\textwidth]{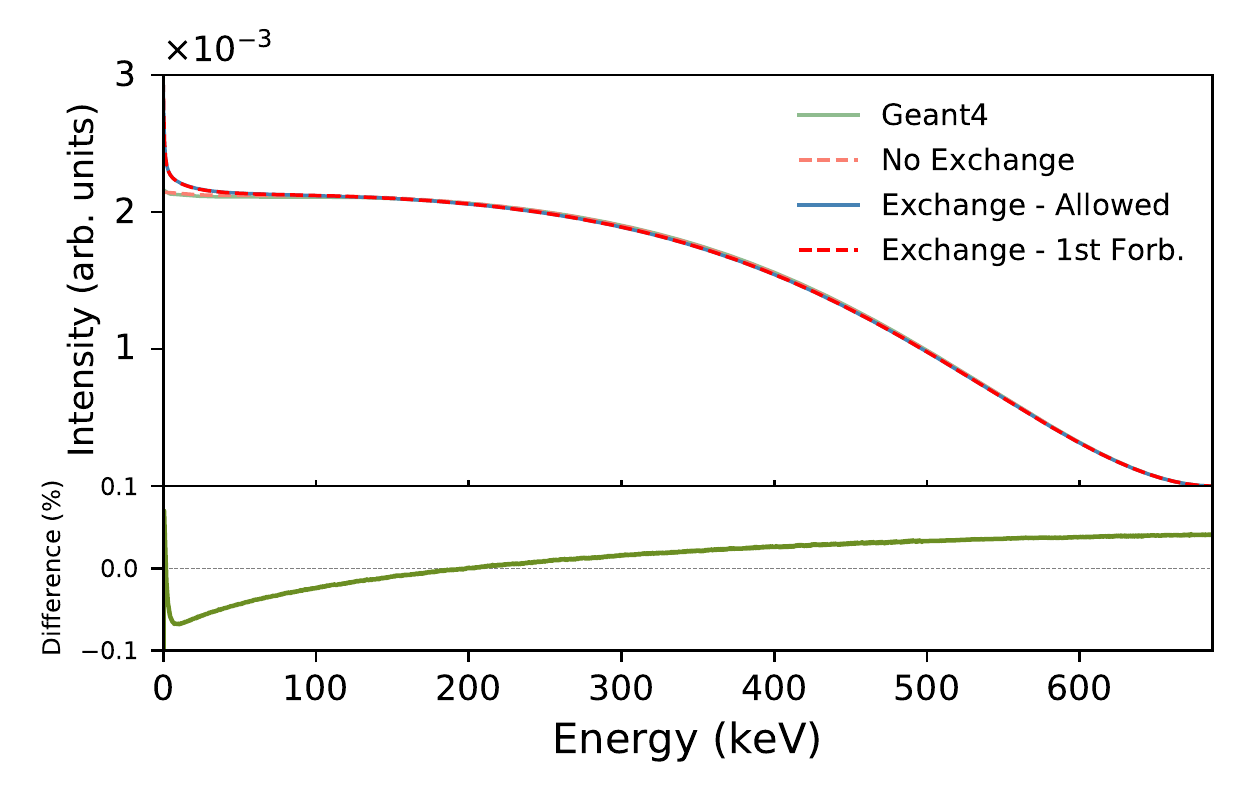}
\end{subfigure}
\begin{subfigure}{0.49\textwidth}
\includegraphics[width=\textwidth]{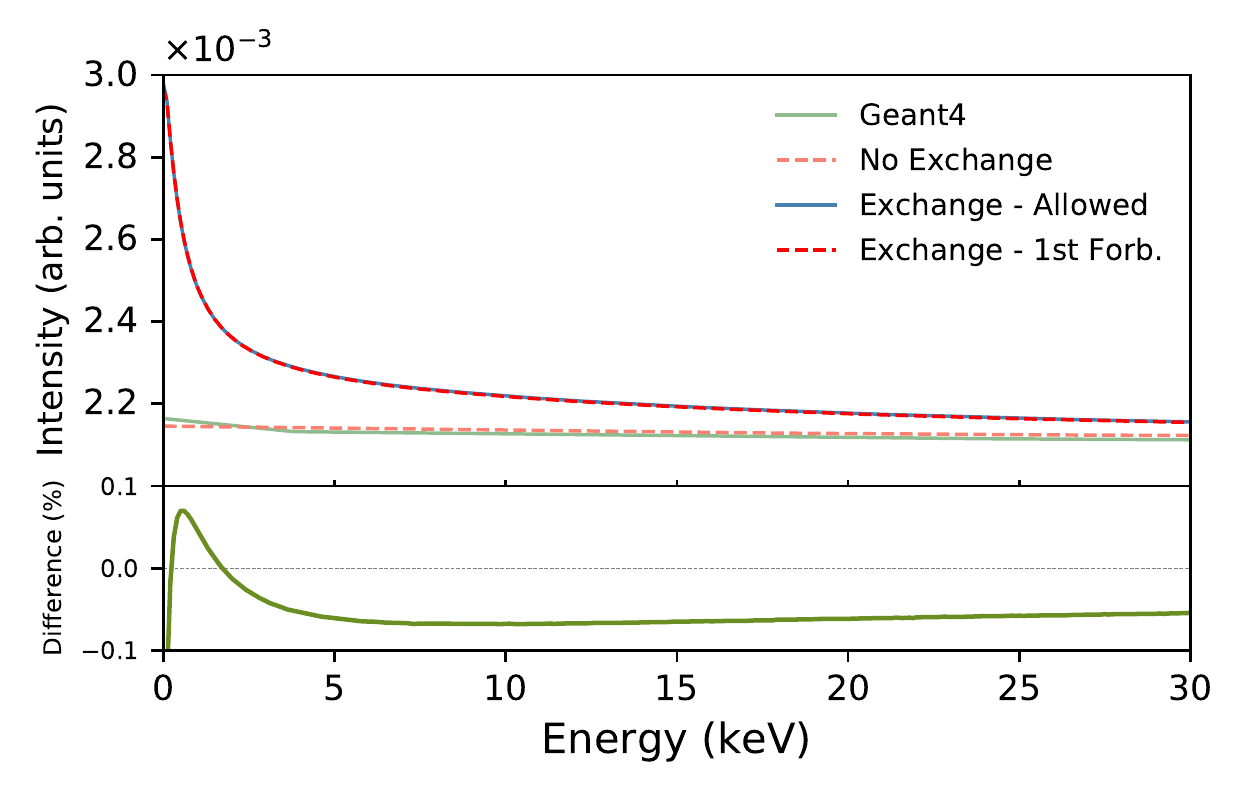}
\end{subfigure}
\caption{Comparison of $\beta$ spectra for the ground-state decay of \KreF shown over the full energy range (left) and at low energies (right). The dashed red line shows the spectrum with the exchange effect calculated using the extended formalism for first-forbidden unique transitions. The lower portion of each figure gives the difference between the spectra with the exchange effect calculated as an allowed and a first-forbidden unique transition. Spectra are normalized over the full energy range. \label{fig:kr85}}
\end{figure*}

The principle result of this work is to show the impact of the full nuclear calculation on the spectral shape. Thus we compare our \Pbtof and \Pbtot spectra to those used in the background model of Ref.~\cite{Aprile:2020tmw}, shown here in blue. As only rank-1 operators contribute in these transitions, $\beta$ electrons can only be created with $\kappa = \pm 1$, i.e. the atomic exchange correction as for an allowed transition in Eq.~(\ref{eq:excorr_all}) is a good approximation. However, the spectra were calculated as allowed in Ref.~\cite{Aprile:2020tmw}, without any nuclear shape adjustment, unlike the present work. 

For both Pb isotopes the present calculations result in less rate in the low energy region of interest for several new physics searches compared to the allowed calculation. The spectra obtained here with $g_{\rm A}=0.85$ predict a 19.0\% and 15.5\% lower event rate in a 1--15~keV energy window from \Pbtof and \Pbtot, respectively. Over the 0.7--1.0 range of $g_{\rm A}$ the corresponding ratios are 15.0--23.2\% and 12.1--19.0\%. For further reference we also show in green the $\beta$ shape generated by the \textsc{GEANT4}~\cite{agostinelli:2002hh} toolkit commonly used to predict background rates and energy spectra in LXe TPC experiments. A detailed description of the $\beta$ spectrum model used in \textsc{GEANT4} can be found in the appendix of~Ref.~\cite{Aprile:2020tmw}.

For an analysis such as that in~\cite{Aprile:2020tmw} performed in a restricted energy window well below the $\beta$ endpoint, differences in the assumed $\beta$ spectrum introduce a background systematic in the lowest energy region used to search for possible new physics signals. To illustrate the size of this systematic we normalized the area under our spectra to that under the allowed shape in a 1--210~keV energy window like that used in~\cite{Aprile:2020tmw}. For the $\beta$ decay from \Pbtof our spectra predict 4.3\%, 5.5\%, and 6.7\% less rate in a 1--15~keV window corresponding to the values $g_{\rm A}=0.7,0.85,1.0$. For \Pbtot these ratios are 6.6\%, 8.4\%, and 10.3\%. Interestingly, these results suggest that the size of the excess observed by XENON1T could in fact be larger than what is reported.

For the first time, the ground state \KreF $\beta$ spectrum has been calculated with the correct atomic exchange effect for this first-forbidden unique transition. Fig.~\ref{fig:kr85} shows our result (red dashed line) compared with three other calculations. The solid blue spectrum is the result given in Ref.~\cite{Aprile:2020tmw}, calculated as a first-forbidden unique transition with an atomic exchange correction as for an allowed transition. The solid green spectrum comes from the model used by \textsc{GEANT4}, and the dashed orange spectrum does not include any exchange correction. The difference between the spectra from this work and Ref.~\cite{Aprile:2020tmw} is given in the lower portion of the figure and is found to be in the range of $\pm0.05$\%. Such a negligible difference comes from a combination of effects. First, seven orbitals contribute to $T_{\pm 1}$ but only four orbitals to $T_{\pm 2}$. Secondly, the exchange correction in Ref.~\cite{Aprile:2020tmw} corresponds to applying the approximation
\begin{equation}
\left[ q^2 + \lambda_2 p^2 \right] \text{ } \longrightarrow \text{ } \approx \left[ q^2 + \lambda_2 p^2 \right] \times (1+\eta_1) \, ,
\end{equation}
which means that we are comparing $\eta_1$ with $\eta_2$, two quantities of similar magnitude. Lastly, the magnitude of the exchange correction factors is most important at low energy and as can be seen from Eqs.~(\ref{eq:excorr_1}) and (\ref{eq:excorr_2}): the energy dependence of the shape factor enhances the influence of $(1+\eta_1)$ and at the opposite reduces the influence of $(1+\eta_2)$. This explains why our extended calculation of the exchange effect in \KreF decay gives a $\beta$ spectrum very close to the approximate spectrum of Ref.~\cite{Aprile:2020tmw}. One can expect similar behavior in every first-forbidden unique transition as long as the transition is not dominated by an accidental cancellation of the NMEs.

\section{\label{sec:conclusion}CONCLUSION}

We have presented improved energy spectra for the ground-state $\beta$ decays of \Pbtof, \Pbtot, and \KreF. Combinations of these three decays form the most significant sources of background in current and future LXe dark matter experiments at low energy, \Pbtof being the most salient of the three. The spectra derived here make use of a nuclear shell model formalism to calculate the relevant NMEs and include corrections for the atomic exchange effect.

We find that the ground-state spectra depend on the weak axial vector coupling $g_{\rm A}$ and therefore produce spectra using a suitable range for its value. Our results predict a 19.0\% and 15.5\% downward shift in background rate from \Pbtof and \Pbtot in the energy region of interest for new physics searches relative to previous predictions. Our assessment of a suitable range for $g_{\rm A}$ suggests that these shifts have an uncertainty of roughly 4\%. The overall impact of nuclear structure effects on the $^{212,214}$Pb spectra is found to be more significant than that from the atomic exchange effect considered previously. An extension of the atomic exchange correction to include first-forbidden unique transitions was applied to the ground-state decay of \KreF. The final spectrum shows very minor differences relative to previous calculations which use an allowed approximation for the exchange effect.

Given the large discrepancy between the present calculation and the allowed approximation in \Pbtof and \Pbtot used previously, a direct measurement of these transitions would be prudent for the reduction of systematic errors in future experimental endeavors. To our knowledge, no direct experimental data for these spectra exist in the lowest energy region of concern for modern experiments. Historical investigations of the $\beta$ spectra from \Pbtof have focused on lines from internal conversion electrons and not on the continuous spectrum below $\sim700$~keV from decays to the $^{214}$Bi ground state. A dedicated measurement of these shapes could, for example, be comprised of a central, low-threshold detector containing a $^{212,214}$Pb source which is surrounded by a highly efficient $\gamma$-ray veto detector. In this configuration, the ground-state decays are reconstructed from the sample of events with a signature in the central detector, but with no coincident signal in the outer veto.

\section{ACKNOWLEDGMENTS}
We thank Harry Nelson for helpful discussions. This work was supported by the U.S. Department of Energy Office of Science under contract number DE-AC02-05CH11231 and by the Academy of Finland under the Academy project no. 318043. J. K. acknowledges the financial support from the Jenny and Antti Wihuri Foundation.

\FloatBarrier
\bibliographystyle{apsrev4-2}
\bibliography{beta}

\begin{thebibliography}{33}%
\makeatletter
\providecommand \@ifxundefined [1]{%
 \@ifx{#1\undefined}
}%
\providecommand \@ifnum [1]{%
 \ifnum #1\expandafter \@firstoftwo
 \else \expandafter \@secondoftwo
 \fi
}%
\providecommand \@ifx [1]{%
 \ifx #1\expandafter \@firstoftwo
 \else \expandafter \@secondoftwo
 \fi
}%
\providecommand \natexlab [1]{#1}%
\providecommand \enquote  [1]{``#1''}%
\providecommand \bibnamefont  [1]{#1}%
\providecommand \bibfnamefont [1]{#1}%
\providecommand \citenamefont [1]{#1}%
\providecommand \href@noop [0]{\@secondoftwo}%
\providecommand \href [0]{\begingroup \@sanitize@url \@href}%
\providecommand \@href[1]{\@@startlink{#1}\@@href}%
\providecommand \@@href[1]{\endgroup#1\@@endlink}%
\providecommand \@sanitize@url [0]{\catcode `\\12\catcode `\$12\catcode
  `\&12\catcode `\#12\catcode `\^12\catcode `\_12\catcode `\%12\relax}%
\providecommand \@@startlink[1]{}%
\providecommand \@@endlink[0]{}%
\providecommand \url  [0]{\begingroup\@sanitize@url \@url }%
\providecommand \@url [1]{\endgroup\@href {#1}{\urlprefix }}%
\providecommand \urlprefix  [0]{URL }%
\providecommand \Eprint [0]{\href }%
\providecommand \doibase [0]{http://dx.doi.org/}%
\providecommand \selectlanguage [0]{\@gobble}%
\providecommand \bibinfo  [0]{\@secondoftwo}%
\providecommand \bibfield  [0]{\@secondoftwo}%
\providecommand \translation [1]{[#1]}%
\providecommand \BibitemOpen [0]{}%
\providecommand \bibitemStop [0]{}%
\providecommand \bibitemNoStop [0]{.\EOS\space}%
\providecommand \EOS [0]{\spacefactor3000\relax}%
\providecommand \BibitemShut  [1]{\csname bibitem#1\endcsname}%
\let\auto@bib@innerbib\@empty
\bibitem [{\citenamefont {Akerib}\ \emph {et~al.}(2018)\citenamefont {Akerib}
  \emph {et~al.}}]{Akerib:2016hcd}%
  \BibitemOpen
  \bibfield  {author} {\bibinfo {author} {\bibfnamefont {D.}~\bibnamefont
  {Akerib}} \emph {et~al.} (\bibinfo {collaboration} {LUX}),\ }\href {\doibase
  10.1016/j.astropartphys.2017.10.014} {\bibfield  {journal} {\bibinfo
  {journal} {Astropart. Phys.}\ }\textbf {\bibinfo {volume} {97}},\ \bibinfo
  {pages} {80} (\bibinfo {year} {2018})},\ \Eprint
  {http://arxiv.org/abs/1605.03844}{arXiv:1605.03844
  [physics.ins-det]}\BibitemShut {NoStop}%
\bibitem [{\citenamefont {Aprile}\ \emph {et~al.}(2017)\citenamefont {Aprile}
  \emph {et~al.}}]{Aprile:2016xhi}%
  \BibitemOpen
  \bibfield  {author} {\bibinfo {author} {\bibfnamefont {E.}~\bibnamefont
  {Aprile}} \emph {et~al.} (\bibinfo {collaboration} {XENON}),\ }\href
  {\doibase 10.1140/epjc/s10052-017-4757-1} {\bibfield  {journal} {\bibinfo
  {journal} {Eur. Phys. J. C}\ }\textbf {\bibinfo {volume} {77}},\ \bibinfo
  {pages} {275} (\bibinfo {year} {2017})},\ \Eprint
  {http://arxiv.org/abs/1612.04284}{arXiv:1612.04284
  [physics.ins-det]}\BibitemShut {NoStop}%
\bibitem [{\citenamefont {Wang}\ \emph {et~al.}(2017)\citenamefont {Wang},
  \citenamefont {Audi}, \citenamefont {Kondev}, \citenamefont {Huang},
  \citenamefont {Naimi},\ and\ \citenamefont {Xu}}]{Wan17}%
  \BibitemOpen
  \bibfield  {author} {\bibinfo {author} {\bibfnamefont {M.}~\bibnamefont
  {Wang}}, \bibinfo {author} {\bibfnamefont {G.}~\bibnamefont {Audi}}, \bibinfo
  {author} {\bibfnamefont {F.}~\bibnamefont {Kondev}}, \bibinfo {author}
  {\bibfnamefont {W.}~\bibnamefont {Huang}}, \bibinfo {author} {\bibfnamefont
  {S.}~\bibnamefont {Naimi}}, \ and\ \bibinfo {author} {\bibfnamefont
  {X.}~\bibnamefont {Xu}},\ }\href {\doibase 10.1088/1674-1137/41/3/030003}
  {\bibfield  {journal} {\bibinfo  {journal} {Chin. Phys. C}\ }\textbf
  {\bibinfo {volume} {41}},\ \bibinfo {pages} {030003} (\bibinfo {year}
  {2017})}\BibitemShut {NoStop}%
\bibitem [{\citenamefont {Wu}(2009)}]{A214datasheet}%
  \BibitemOpen
  \bibfield  {author} {\bibinfo {author} {\bibfnamefont {S.-C.}\ \bibnamefont
  {Wu}},\ }\href {\doibase https://doi.org/10.1016/j.nds.2009.02.002}
  {\bibfield  {journal} {\bibinfo  {journal} {Nuclear Data Sheets}\ }\textbf
  {\bibinfo {volume} {110}},\ \bibinfo {pages} {681 } (\bibinfo {year}
  {2009})}\BibitemShut {NoStop}%
\bibitem [{\citenamefont {Browne}(2005)}]{A212datasheet}%
  \BibitemOpen
  \bibfield  {author} {\bibinfo {author} {\bibfnamefont {E.}~\bibnamefont
  {Browne}},\ }\href {\doibase https://doi.org/10.1016/j.nds.2005.01.002}
  {\bibfield  {journal} {\bibinfo  {journal} {Nuclear Data Sheets}\ }\textbf
  {\bibinfo {volume} {104}},\ \bibinfo {pages} {427 } (\bibinfo {year}
  {2005})}\BibitemShut {NoStop}%
\bibitem [{\citenamefont {Singh}\ and\ \citenamefont
  {Chen}(2014)}]{A85datasheet}%
  \BibitemOpen
  \bibfield  {author} {\bibinfo {author} {\bibfnamefont {B.}~\bibnamefont
  {Singh}}\ and\ \bibinfo {author} {\bibfnamefont {J.}~\bibnamefont {Chen}},\
  }\href {\doibase https://doi.org/10.1016/j.nds.2014.01.001} {\bibfield
  {journal} {\bibinfo  {journal} {Nuclear Data Sheets}\ }\textbf {\bibinfo
  {volume} {116}},\ \bibinfo {pages} {1 } (\bibinfo {year} {2014})}\BibitemShut
  {NoStop}%
\bibitem [{\citenamefont {Akerib}\ \emph {et~al.}(2020)\citenamefont {Akerib}
  \emph {et~al.}}]{LZ:2018}%
  \BibitemOpen
  \bibfield  {author} {\bibinfo {author} {\bibfnamefont {D.}~\bibnamefont
  {Akerib}} \emph {et~al.} (\bibinfo {collaboration} {LUX-ZEPLIN}),\ }\href
  {\doibase 10.1103/PhysRevD.101.052002} {\bibfield  {journal} {\bibinfo
  {journal} {Phys. Rev. D}\ }\textbf {\bibinfo {volume} {101}},\ \bibinfo
  {pages} {052002} (\bibinfo {year} {2020})},\ \Eprint
  {http://arxiv.org/abs/1802.06039}{arXiv:1802.06039 [astro-ph.IM]}\BibitemShut
  {NoStop}%
\bibitem [{\citenamefont {Aprile}\ \emph
  {et~al.}(2020{\natexlab{a}})\citenamefont {Aprile} \emph {et~al.}}]{xntSens}%
  \BibitemOpen
  \bibfield  {author} {\bibinfo {author} {\bibfnamefont {E.}~\bibnamefont
  {Aprile}} \emph {et~al.} (\bibinfo {collaboration} {XENON}),\ }\href@noop {}
  {\  (\bibinfo {year} {2020}{\natexlab{a}})},\ \Eprint
  {http://arxiv.org/abs/2007.08796}{arXiv:2007.08796
  [physics.ins-det]}\BibitemShut {NoStop}%
\bibitem [{\citenamefont {Aprile}\ \emph
  {et~al.}(2020{\natexlab{b}})\citenamefont {Aprile} \emph
  {et~al.}}]{Aprile:2020tmw}%
  \BibitemOpen
  \bibfield  {author} {\bibinfo {author} {\bibfnamefont {E.}~\bibnamefont
  {Aprile}} \emph {et~al.} (\bibinfo {collaboration} {XENON Collaboration}),\
  }\href {\doibase 10.1103/PhysRevD.102.072004} {\bibfield  {journal} {\bibinfo
   {journal} {Phys. Rev. D}\ }\textbf {\bibinfo {volume} {102}},\ \bibinfo
  {pages} {072004} (\bibinfo {year} {2020}{\natexlab{b}})}\BibitemShut
  {NoStop}%
\bibitem [{\citenamefont {Mustonen}\ \emph {et~al.}(2006)\citenamefont
  {Mustonen}, \citenamefont {Aunola},\ and\ \citenamefont
  {Suhonen}}]{Mustonen2006}%
  \BibitemOpen
  \bibfield  {author} {\bibinfo {author} {\bibfnamefont {M.~T.}\ \bibnamefont
  {Mustonen}}, \bibinfo {author} {\bibfnamefont {M.}~\bibnamefont {Aunola}}, \
  and\ \bibinfo {author} {\bibfnamefont {J.}~\bibnamefont {Suhonen}},\
  }\href@noop {} {\bibfield  {journal} {\bibinfo  {journal} {Phys. Rev. C}\
  }\textbf {\bibinfo {volume} {73}},\ \bibinfo {pages} {054301} (\bibinfo
  {year} {2006})}\BibitemShut {NoStop}%
\bibitem [{\citenamefont {Haaranen}\ \emph {et~al.}(2016)\citenamefont
  {Haaranen}, \citenamefont {Srivastava},\ and\ \citenamefont
  {Suhonen}}]{Haaranen2016}%
  \BibitemOpen
  \bibfield  {author} {\bibinfo {author} {\bibfnamefont {M.}~\bibnamefont
  {Haaranen}}, \bibinfo {author} {\bibfnamefont {P.~C.}\ \bibnamefont
  {Srivastava}}, \ and\ \bibinfo {author} {\bibfnamefont {J.}~\bibnamefont
  {Suhonen}},\ }\href@noop {} {\bibfield  {journal} {\bibinfo  {journal} {Phys.
  Rev. C}\ }\textbf {\bibinfo {volume} {93}},\ \bibinfo {pages} {034308}
  (\bibinfo {year} {2016})}\BibitemShut {NoStop}%
\bibitem [{\citenamefont {Haaranen}\ \emph {et~al.}(2017)\citenamefont
  {Haaranen}, \citenamefont {Kotila},\ and\ \citenamefont
  {Suhonen}}]{Haaranen2017}%
  \BibitemOpen
  \bibfield  {author} {\bibinfo {author} {\bibfnamefont {M.}~\bibnamefont
  {Haaranen}}, \bibinfo {author} {\bibfnamefont {J.}~\bibnamefont {Kotila}}, \
  and\ \bibinfo {author} {\bibfnamefont {J.}~\bibnamefont {Suhonen}},\
  }\href@noop {} {\bibfield  {journal} {\bibinfo  {journal} {Phys. Rev. C}\
  }\textbf {\bibinfo {volume} {95}},\ \bibinfo {pages} {024327} (\bibinfo
  {year} {2017})}\BibitemShut {NoStop}%
\bibitem [{\citenamefont {Kostensalo}\ and\ \citenamefont
  {Suhonen}(2017)}]{Kostensalo2017}%
  \BibitemOpen
  \bibfield  {author} {\bibinfo {author} {\bibfnamefont {J.}~\bibnamefont
  {Kostensalo}}\ and\ \bibinfo {author} {\bibfnamefont {J.}~\bibnamefont
  {Suhonen}},\ }\href@noop {} {\bibfield  {journal} {\bibinfo  {journal} {Phys.
  Rev. C}\ }\textbf {\bibinfo {volume} {96}},\ \bibinfo {pages} {024317}
  (\bibinfo {year} {2017})}\BibitemShut {NoStop}%
\bibitem [{\citenamefont {Ejiri}\ \emph {et~al.}(2019)\citenamefont {Ejiri},
  \citenamefont {Suhonen},\ and\ \citenamefont {Zuber}}]{Ejiri2019}%
  \BibitemOpen
  \bibfield  {author} {\bibinfo {author} {\bibfnamefont {H.}~\bibnamefont
  {Ejiri}}, \bibinfo {author} {\bibfnamefont {J.}~\bibnamefont {Suhonen}}, \
  and\ \bibinfo {author} {\bibfnamefont {K.}~\bibnamefont {Zuber}},\
  }\href@noop {} {\bibfield  {journal} {\bibinfo  {journal} {Phys. Rep.}\
  }\textbf {\bibinfo {volume} {797}},\ \bibinfo {pages} {1} (\bibinfo {year}
  {2019})}\BibitemShut {NoStop}%
\bibitem [{\citenamefont {Kumar}\ \emph {et~al.}(2020)\citenamefont {Kumar},
  \citenamefont {Srivastava}, \citenamefont {Kostensalo},\ and\ \citenamefont
  {Suhonen}}]{Kumar2020}%
  \BibitemOpen
  \bibfield  {author} {\bibinfo {author} {\bibfnamefont {A.}~\bibnamefont
  {Kumar}}, \bibinfo {author} {\bibfnamefont {P.~C.}\ \bibnamefont
  {Srivastava}}, \bibinfo {author} {\bibfnamefont {J.}~\bibnamefont
  {Kostensalo}}, \ and\ \bibinfo {author} {\bibfnamefont {J.}~\bibnamefont
  {Suhonen}},\ }\href {\doibase 10.1103/PhysRevC.101.064304} {\bibfield
  {journal} {\bibinfo  {journal} {Phys. Rev. C}\ }\textbf {\bibinfo {volume}
  {101}},\ \bibinfo {pages} {064304} (\bibinfo {year} {2020})}\BibitemShut
  {NoStop}%
\bibitem [{\citenamefont {Hardy}\ \emph {et~al.}(1990)\citenamefont {Hardy},
  \citenamefont {Towner}, \citenamefont {Koslowsky}, \citenamefont {Hagberg},\
  and\ \citenamefont {Schmeing}}]{Hardy1990}%
  \BibitemOpen
  \bibfield  {author} {\bibinfo {author} {\bibfnamefont {J.}~\bibnamefont
  {Hardy}}, \bibinfo {author} {\bibfnamefont {I.}~\bibnamefont {Towner}},
  \bibinfo {author} {\bibfnamefont {V.}~\bibnamefont {Koslowsky}}, \bibinfo
  {author} {\bibfnamefont {E.}~\bibnamefont {Hagberg}}, \ and\ \bibinfo
  {author} {\bibfnamefont {H.}~\bibnamefont {Schmeing}},\ }\href {\doibase
  https://doi.org/10.1016/0375-9474(90)90086-2} {\bibfield  {journal} {\bibinfo
   {journal} {Nuclear Physics A}\ }\textbf {\bibinfo {volume} {509}},\ \bibinfo
  {pages} {429 } (\bibinfo {year} {1990})}\BibitemShut {NoStop}%
\bibitem [{\citenamefont {Behrens}\ and\ \citenamefont
  {B{\"{u}}hring}(1982)}]{Behrens1982}%
  \BibitemOpen
  \bibfield  {author} {\bibinfo {author} {\bibfnamefont {H.}~\bibnamefont
  {Behrens}}\ and\ \bibinfo {author} {\bibfnamefont {W.}~\bibnamefont
  {B{\"{u}}hring}},\ }\href@noop {} {\emph {\bibinfo {title} {Electron Radial
  Wave Functions and Nuclear Beta Decay}}}\ (\bibinfo  {publisher} {Clarendon,
  Oxford},\ \bibinfo {year} {1982})\BibitemShut {NoStop}%
\bibitem [{\citenamefont {Brown}\ and\ \citenamefont {Rae}(2014)}]{nushellx}%
  \BibitemOpen
  \bibfield  {author} {\bibinfo {author} {\bibfnamefont {B.~A.}\ \bibnamefont
  {Brown}}\ and\ \bibinfo {author} {\bibfnamefont {W.~D.~M.}\ \bibnamefont
  {Rae}},\ }\href@noop {} {\bibfield  {journal} {\bibinfo  {journal} {Nuclear
  Data Sheets}\ }\textbf {\bibinfo {volume} {120}},\ \bibinfo {pages} {115}
  (\bibinfo {year} {2014})}\BibitemShut {NoStop}%
\bibitem [{\citenamefont {Honma}\ \emph {et~al.}(2009)\citenamefont {Honma},
  \citenamefont {Otsuka}, \citenamefont {Mizusaki},\ and\ \citenamefont
  {Hjorth-Jensen}}]{Honma2009}%
  \BibitemOpen
  \bibfield  {author} {\bibinfo {author} {\bibfnamefont {M.}~\bibnamefont
  {Honma}}, \bibinfo {author} {\bibfnamefont {T.}~\bibnamefont {Otsuka}},
  \bibinfo {author} {\bibfnamefont {T.}~\bibnamefont {Mizusaki}}, \ and\
  \bibinfo {author} {\bibfnamefont {M.}~\bibnamefont {Hjorth-Jensen}},\ }\href
  {\doibase 10.1103/PhysRevC.80.064323} {\bibfield  {journal} {\bibinfo
  {journal} {Phys. Rev. C}\ }\textbf {\bibinfo {volume} {80}},\ \bibinfo
  {pages} {064323} (\bibinfo {year} {2009})}\BibitemShut {NoStop}%
\bibitem [{\citenamefont {Warburton}\ and\ \citenamefont
  {Brown}(1991)}]{Warburton91}%
  \BibitemOpen
  \bibfield  {author} {\bibinfo {author} {\bibfnamefont {E.~K.}\ \bibnamefont
  {Warburton}}\ and\ \bibinfo {author} {\bibfnamefont {B.~A.}\ \bibnamefont
  {Brown}},\ }\href {\doibase 10.1103/PhysRevC.43.602} {\bibfield  {journal}
  {\bibinfo  {journal} {Phys. Rev. C}\ }\textbf {\bibinfo {volume} {43}},\
  \bibinfo {pages} {602} (\bibinfo {year} {1991})}\BibitemShut {NoStop}%
\bibitem [{\citenamefont {Suhonen}(2017)}]{Suhonen2017}%
  \BibitemOpen
  \bibfield  {author} {\bibinfo {author} {\bibfnamefont {J.~T.}\ \bibnamefont
  {Suhonen}},\ }\href {\doibase 10.3389/fphy.2017.00055} {\bibfield  {journal}
  {\bibinfo  {journal} {Frontiers in Physics}\ }\textbf {\bibinfo {volume}
  {5}},\ \bibinfo {pages} {55} (\bibinfo {year} {2017})}\BibitemShut {NoStop}%
\bibitem [{\citenamefont {Zhi}\ \emph {et~al.}(2013)\citenamefont {Zhi},
  \citenamefont {Caurier}, \citenamefont {Cuenca-Garc\'{\i}a}, \citenamefont
  {Langanke}, \citenamefont {Mart\'{\i}nez-Pinedo},\ and\ \citenamefont
  {Sieja}}]{Zhi2013}%
  \BibitemOpen
  \bibfield  {author} {\bibinfo {author} {\bibfnamefont {Q.}~\bibnamefont
  {Zhi}}, \bibinfo {author} {\bibfnamefont {E.}~\bibnamefont {Caurier}},
  \bibinfo {author} {\bibfnamefont {J.~J.}\ \bibnamefont {Cuenca-Garc\'{\i}a}},
  \bibinfo {author} {\bibfnamefont {K.}~\bibnamefont {Langanke}}, \bibinfo
  {author} {\bibfnamefont {G.}~\bibnamefont {Mart\'{\i}nez-Pinedo}}, \ and\
  \bibinfo {author} {\bibfnamefont {K.}~\bibnamefont {Sieja}},\ }\href
  {\doibase 10.1103/PhysRevC.87.025803} {\bibfield  {journal} {\bibinfo
  {journal} {Phys. Rev. C}\ }\textbf {\bibinfo {volume} {87}},\ \bibinfo
  {pages} {025803} (\bibinfo {year} {2013})}\BibitemShut {NoStop}%
\bibitem [{\citenamefont {Al~Kharusi}\ \emph {et~al.}(2020)\citenamefont
  {Al~Kharusi} \emph {et~al.}}]{EXO-200}%
  \BibitemOpen
  \bibfield  {author} {\bibinfo {author} {\bibfnamefont {S.}~\bibnamefont
  {Al~Kharusi}} \emph {et~al.} (\bibinfo {collaboration} {EXO-200}),\
  }\href@noop {} {\bibfield  {journal} {\bibinfo  {journal} {Phys. Rev. Lett.}\
  }\textbf {\bibinfo {volume} {124}},\ \bibinfo {pages} {232502} (\bibinfo
  {year} {2020})}\BibitemShut {NoStop}%
\bibitem [{\citenamefont {Kostensalo}\ and\ \citenamefont
  {Suhonen}(2018)}]{kostensalo2018}%
  \BibitemOpen
  \bibfield  {author} {\bibinfo {author} {\bibfnamefont {J.}~\bibnamefont
  {Kostensalo}}\ and\ \bibinfo {author} {\bibfnamefont {J.}~\bibnamefont
  {Suhonen}},\ }\href {\doibase 10.1142/S0217751X1843008X} {\bibfield
  {journal} {\bibinfo  {journal} {International Journal of Modern Physics A}\
  }\textbf {\bibinfo {volume} {33}},\ \bibinfo {pages} {1843008} (\bibinfo
  {year} {2018})},\ \Eprint
  {http://arxiv.org/abs/https://doi.org/10.1142/S0217751X1843008X}{https://doi.org/10.1142/S0217751X1843008X}\BibitemShut
  {NoStop}%
\bibitem [{\citenamefont {Hayen}\ \emph {et~al.}(2018)\citenamefont {Hayen},
  \citenamefont {Severijns}, \citenamefont {Bodek}, \citenamefont {Rozpedzik},\
  and\ \citenamefont {Mougeot}}]{Hay18}%
  \BibitemOpen
  \bibfield  {author} {\bibinfo {author} {\bibfnamefont {L.}~\bibnamefont
  {Hayen}}, \bibinfo {author} {\bibfnamefont {N.}~\bibnamefont {Severijns}},
  \bibinfo {author} {\bibfnamefont {K.}~\bibnamefont {Bodek}}, \bibinfo
  {author} {\bibfnamefont {D.}~\bibnamefont {Rozpedzik}}, \ and\ \bibinfo
  {author} {\bibfnamefont {X.}~\bibnamefont {Mougeot}},\ }\href {\doibase
  10.1103/RevModPhys.90.015008} {\bibfield  {journal} {\bibinfo  {journal}
  {Rev. Mod. Phys.}\ }\textbf {\bibinfo {volume} {90}},\ \bibinfo {pages}
  {015008} (\bibinfo {year} {2018})}\BibitemShut {NoStop}%
\bibitem [{\citenamefont {Harston}\ and\ \citenamefont
  {Pyper}(1992)}]{Harston92}%
  \BibitemOpen
  \bibfield  {author} {\bibinfo {author} {\bibfnamefont {M.~R.}\ \bibnamefont
  {Harston}}\ and\ \bibinfo {author} {\bibfnamefont {N.~C.}\ \bibnamefont
  {Pyper}},\ }\href {\doibase 10.1103/PhysRevA.45.6282} {\bibfield  {journal}
  {\bibinfo  {journal} {Phys. Rev. A}\ }\textbf {\bibinfo {volume} {45}},\
  \bibinfo {pages} {6282} (\bibinfo {year} {1992})}\BibitemShut {NoStop}%
\bibitem [{\citenamefont {Mougeot}\ and\ \citenamefont {Bisch}(2014)}]{Mou14}%
  \BibitemOpen
  \bibfield  {author} {\bibinfo {author} {\bibfnamefont {X.}~\bibnamefont
  {Mougeot}}\ and\ \bibinfo {author} {\bibfnamefont {C.}~\bibnamefont
  {Bisch}},\ }\href {\doibase 10.1103/PhysRevA.90.012501} {\bibfield  {journal}
  {\bibinfo  {journal} {Phys. Rev. A}\ }\textbf {\bibinfo {volume} {90}},\
  \bibinfo {pages} {012501} (\bibinfo {year} {2014})}\BibitemShut {NoStop}%
\bibitem [{\citenamefont {Kossert}\ and\ \citenamefont
  {Mougeot}(2015)}]{Kossert15}%
  \BibitemOpen
  \bibfield  {author} {\bibinfo {author} {\bibfnamefont {K.}~\bibnamefont
  {Kossert}}\ and\ \bibinfo {author} {\bibfnamefont {X.}~\bibnamefont
  {Mougeot}},\ }\href {\doibase https://doi.org/10.1016/j.apradiso.2015.03.017}
  {\bibfield  {journal} {\bibinfo  {journal} {Appl. Radiat. Isot.}\ }\textbf
  {\bibinfo {volume} {101}},\ \bibinfo {pages} {40} (\bibinfo {year}
  {2015})}\BibitemShut {NoStop}%
\bibitem [{\citenamefont {Kossert}\ \emph {et~al.}(2018)\citenamefont
  {Kossert}, \citenamefont {Marganiec-Galazka}, \citenamefont {Mougeot},\ and\
  \citenamefont {N{\"a}hle}}]{Kossert18}%
  \BibitemOpen
  \bibfield  {author} {\bibinfo {author} {\bibfnamefont {K.}~\bibnamefont
  {Kossert}}, \bibinfo {author} {\bibfnamefont {J.}~\bibnamefont
  {Marganiec-Galazka}}, \bibinfo {author} {\bibfnamefont {X.}~\bibnamefont
  {Mougeot}}, \ and\ \bibinfo {author} {\bibfnamefont {O.}~\bibnamefont
  {N{\"a}hle}},\ }\href {\doibase
  https://doi.org/10.1016/j.apradiso.2017.06.015} {\bibfield  {journal}
  {\bibinfo  {journal} {Appl. Radiat. Isot.}\ }\textbf {\bibinfo {volume}
  {134}},\ \bibinfo {pages} {212} (\bibinfo {year} {2018})}\BibitemShut
  {NoStop}%
\bibitem [{\citenamefont {Pyper}\ and\ \citenamefont
  {Harston}(1988)}]{Pyper88}%
  \BibitemOpen
  \bibfield  {author} {\bibinfo {author} {\bibfnamefont {N.~C.}\ \bibnamefont
  {Pyper}}\ and\ \bibinfo {author} {\bibfnamefont {M.~R.}\ \bibnamefont
  {Harston}},\ }\href {\doibase https://doi.org/10.1098/rspa.1988.0128}
  {\bibfield  {journal} {\bibinfo  {journal} {Proc. Roy. Soc. Lond. A}\
  }\textbf {\bibinfo {volume} {420}},\ \bibinfo {pages} {277} (\bibinfo {year}
  {1988})}\BibitemShut {NoStop}%
\bibitem [{\citenamefont {Towner}\ and\ \citenamefont {Hardy}(2008)}]{Tow08}%
  \BibitemOpen
  \bibfield  {author} {\bibinfo {author} {\bibfnamefont {I.~S.}\ \bibnamefont
  {Towner}}\ and\ \bibinfo {author} {\bibfnamefont {J.~C.}\ \bibnamefont
  {Hardy}},\ }\href {\doibase 10.1103/PhysRevC.77.025501} {\bibfield  {journal}
  {\bibinfo  {journal} {Phys. Rev. C}\ }\textbf {\bibinfo {volume} {77}},\
  \bibinfo {pages} {025501} (\bibinfo {year} {2008})}\BibitemShut {NoStop}%
\bibitem [{\citenamefont {Kotochigova}\ \emph {et~al.}(1997)\citenamefont
  {Kotochigova}, \citenamefont {Levine}, \citenamefont {Shirley}, \citenamefont
  {Stiles},\ and\ \citenamefont {Clark}}]{Kot97}%
  \BibitemOpen
  \bibfield  {author} {\bibinfo {author} {\bibfnamefont {S.}~\bibnamefont
  {Kotochigova}}, \bibinfo {author} {\bibfnamefont {Z.~H.}\ \bibnamefont
  {Levine}}, \bibinfo {author} {\bibfnamefont {E.~L.}\ \bibnamefont {Shirley}},
  \bibinfo {author} {\bibfnamefont {M.~D.}\ \bibnamefont {Stiles}}, \ and\
  \bibinfo {author} {\bibfnamefont {C.~W.}\ \bibnamefont {Clark}},\ }\href
  {\doibase 10.1103/PhysRevA.55.191} {\bibfield  {journal} {\bibinfo  {journal}
  {Phys. Rev. A}\ }\textbf {\bibinfo {volume} {55}},\ \bibinfo {pages} {191}
  (\bibinfo {year} {1997})}\BibitemShut {NoStop}%
\bibitem [{\citenamefont {Agostinelli}\ \emph {et~al.}(2003)\citenamefont
  {Agostinelli} \emph {et~al.}}]{agostinelli:2002hh}%
  \BibitemOpen
  \bibfield  {author} {\bibinfo {author} {\bibfnamefont {S.}~\bibnamefont
  {Agostinelli}} \emph {et~al.} (\bibinfo {collaboration}
  {\href{http://www.geant4.org/geant4/}{GEANT4}}),\ }\href {\doibase
  10.1016/S0168-9002(03)01368-8} {\bibfield  {journal} {\bibinfo  {journal}
  {Nucl. Instrum. Meth.}\ }\textbf {\bibinfo {volume} {A506}},\ \bibinfo
  {pages} {250} (\bibinfo {year} {2003})}\BibitemShut {NoStop}%
\end{thebibliography}%

\end{document}